\documentclass[twocolumn,aps,superscriptaddress,showpacs,nofootinbib,floatfix]{revtex4}
\usepackage{listings}
\usepackage[utf8]{inputenc}
\usepackage{hyperref}
\usepackage{amsmath}
\usepackage{amssymb}
\usepackage{amsfonts}
\usepackage{tabularx}
\usepackage{booktabs}
\usepackage{graphicx}
\usepackage{color}
\usepackage{multirow}
\usepackage[inline]{enumitem}
\usepackage[T2A,T1]{fontenc}
\graphicspath{{fig/}}
\definecolor{theblue}{RGB}{0,50,230}

\usepackage{hyperref}
\hypersetup{
  colorlinks=true,
  linkcolor=theblue,
  citecolor=theblue,
  urlcolor=theblue
}

\newcommand {\avg}[1]{\ensuremath{\langle\kern-1.0pt\langle#1\rangle\kern-1.0pt\rangle}}

\newlength\cmsFigWidth

\usepackage[normalem]{ulem}  

\renewcommand\sout{\bgroup \color{red} \ULdepth=-.5ex \ULset}

\begin{document}


\title{Centrality and Transverse momentum dependencies of Hadrons in Xe+Xe collisions at $\sqrt{s_{NN}}=5.44$ TeV from a multi-phase transport model}


\author{Lilin Zhu}
\affiliation{Department of Physics, Sichuan University, Chengdu 610064, China}
\author{Hua Zheng}\email{zhengh@snnu.edu.cn}
\affiliation{School of Physics and Information Technology, Shaanxi Normal University, Xi'an 710119, China}
\author{Ruimin Kong}
\affiliation{Chengdu Experimental Foreign Languages School, Chengdu 610064, China }


\begin{abstract}
In this paper, we study and predict the charged-particle pseudorapidity multiplicity density ($\frac{dN_{ch}}{d\eta}$), transverse momentum spectra of identified particles and their ratios in relativistic heavy ion collisions at the Large Hadron Collider (LHC), using the string-melting version of a multi-phase transport (AMPT) model with improved quark coalescence method. Results of the charged-particle pseudorapidity multiplicity density from AMPT model calculations for Pb+Pb collisions at $\sqrt{s_{NN}}=5.02$ TeV are compared with the experimental data. Good agreements are generally found between theoretical calculations and experimental data.  We predict $\frac{dN_{ch}}{d\eta}$ for Xe+Xe collisions at $\sqrt{s_{NN}}=5.44$ TeV at different centralities, and $p_T$ spectra of charged pions, kaons and protons, and their ratios $K/\pi$ and $p/\pi$ in Pb+Pb collisions at $\sqrt{s_{NN}}=5.02$ TeV and Xe+Xe collisions at $\sqrt{s_{NN}}=5.44$ TeV that are being studied at LHC. The $p_T$ spectra of identified particles in Pb+Pb collisions from the improved AMPT model are compared and found to be consistent with results from the iEBE-VISHNU hybrid model with TRENTo initial condition. 

\end{abstract}

\maketitle

\section{introduction}
Among many properties of the dense matter created in the heavy ion collisions at BNL Relativistic Heavy-Ion Collider (RHIC) and CERN Large Hadron Collider (LHC), the particle production is evolving into a mature field and always the subject of particular interest. The final particle production has been studied at RHIC and LHC for different systems, such as Au+Au, Cu+Cu, Pb+Pb and Xe+Xe, at different colliding energies from $\sqrt{s_{NN}}=7.7$ GeV to 5.44 TeV \cite{auau62strange, cucu200pi0phenixadd, auau200pip, Adam:2015kca, Adam:2016ddh, Acharya:2018eaq}. In the past decades, the theoretical investigation of particle production in relativistic heavy ion collisions have been performed extensively with hydrodynamic model \cite{Zhao:2017yhj, Zhao:2018lyf, Zhao:2017rgg}, statistical model \cite{pbm}, transport model \cite{Ma:2016fve, He:2017tla} and coalescence model \cite{vg, rf, hy1}. The models are independent of each other with great success in interpreting the experimental data.

A multi-phase transport (AMPT) model, which was proposed about two decades ago \cite{Zhang:1999bd}, has been widely applied to describe the space-time evolution of relativistic heavy ion collisions and the experimental observables, such as $p_T$ spectra, azimuthal anisotropy and longitudinal correlation of particles, with great success by the theorists and experimentalists \cite{Lin:2004en, Ma:2016fve, He:2017tla}. There are two different versions of AMPT model developed, the default version and the string-melting version, with different model design \cite{Lin:2004en}. The default version of AMPT model can describe the rapidity distributions and transverse momentum spectra of identified particles, but it underestimates the elliptic flow at RHIC \cite{Zhang:1999bd, Lin:2000cx}. For the string-melting version, it can reproduce the charged pions and kaons spectra and elliptic flow in Au+Au collisions at $\sqrt{s_{NN}}=200$ GeV and Pb+Pb collisions at $\sqrt{s_{NN}}=2.76$ TeV \cite{Ma:2016fve}. Unfortunately, the two current versions fail to describe the baryon production \cite{Lin:2004en, Zhu:2015voa}. Therefore, an improved AMPT model with new quark coalescence mechanism in the string-melting version was recently proposed \cite{He:2017tla}.  After modifying the criteria of coalescing partons, the improved AMPT model succeeds to describe the meson and baryon observables in general at RHIC and LHC energies from $\sqrt{s_{NN}}=62.4$ GeV to 2.76 TeV.

In late 2017, Xe+Xe collisions at $\sqrt{s_{NN}}=5.44$ TeV has been carried out at LHC \cite{Acharya:2018eaq, Acharya:2018hhy, Ragoni:2018mqx}. The mass number of xenon being roughly halfway between that of proton and that of a lead nucleus, upcoming data from the Xe+Xe run together with the data from Pb+Pb at $\sqrt{s_{NN}}=5.02$ TeV offer a unique possibility to test the predictive power of the improved AMPT model. The recent theoretical studies for Xe+Xe collisions focus on the azimuthal anisotropy using different models, such as hydrodynamic model \cite{Giacalone:2017dud}, transport model \cite {Tripathy:2018bib} and hybrid model \cite{Shen:2014vra, Eskola:2017bup}. But in this paper, we will focus on the centrality dependence of charged-particle pseudorapidity multiplicity density, transverse momentum spectra of charged pions, kaons and protons and their ratios both in Pb+Pb collisions at $\sqrt{s_{NN}}=5.02$ TeV and Xe+Xe collisions at $\sqrt{s_{NN}}=5.44$ TeV. The comparisons between the model predictions and the experimental data published by the experimental collaborations in the near future will not only examine the validity of the AMPT model, but also shed light on the understanding the space-time evolution of the heavy ion collisions.

The paper is organized as follows. In Section~\ref{ampt}, we briefly review the improved AMPT model used for the present study. Sec.~\ref{results} presents and discusses the charged-particle pseudorapidity multiplicity density, transverse momentum spectra of identified particles and their ratios in Pb+Pb collisions at $\sqrt{s_{NN}}=5.02$ TeV and Xe+Xe collisions at $\sqrt{s_{NN}}=5.44$ TeV, generated with the improved AMPT model. A summary is given in Section~\ref{summary}.

\section{the improved AMPT model}\label{ampt}
The string-melting AMPT model is a hybrid Monte Carlo transport model~\cite{Lin:2004en} consisting of four main components, i.e., the initial conditions, two-body elastic parton cascade, hadronization and hadronic interactions. The initial particle distributions, including spatial and momentum, generated by the heavy ion jet interaction generator (HIJING) model~\cite{Wang:1991hta}. All hadrons produced from string fragmentation in the HIJING model are converted to their valence quarks and antiquarks, which mimics the quark and antiquark plasma produced in relativistic heavy ion collisions.
The parton evolution in time and space is simulated by Zhang's parton cascade (ZPC) model, where the parton scattering cross section is obtained from the perturbative QCD with screening mass. After the parton scatterings, a naive spatial quark coalescence model is adopted to describe the hadronization process. In the current AMPT model, a quark (or an antiquark) always searches a nearby meson partner to form a meson before searching nearby baryon (or antibaryon) partners in coordinate space and ignoring the relative momentum between the coalescing partons, which results in the unreasonable treatment of baryon production. Although some improvements have been introduced in the version of AMPT model used in Ref. \cite{Zhu:2015voa} by changing the coalescence order between mesons and baryons, it still didn't solve the baryon production during hadronization. Recently, the quark coalescence mechanism have been improved in the string-melting AMPT model in Ref. \cite{He:2017tla}. The improved quark coalescence method in AMPT model removes the constraint that forced the quark coalescence process to conserve the numbers of baryons, antibaryons and mesons for each event, but the number of the net-baryons and the number of net strangeness are still conserved for each event. The new quark coalescence method allows a quark to form either a meson or baryon, depending on the distance to its coalescence partner(s). When both its meson partner and baryon partners are available, a new coalescence parameter $r_{BM}$ is introduced to control the relative probability of a quark forming a baryon instead of forming a meson. Within this framework, there is no priority to form mesons or baryons. In this sense, this coalescence picture for hadronization is much more reasonable and physical. Finally, the scatterings among hadrons are described by a relativistic transport (ART) model~\cite{Li:1995pra}, which includes two-particle elastic and inelastic scatterings.   

In the improved string-melting AMPT model with the new quark coalescence method, we adopt the latest version Ampt-v1.31t1-v2.31t1\footnote{This version has not been available online yet. It was provided by the private communication.} from Ref. \cite{He:2017tla} with the Lund string fragmentation parameters $a=0.2$ and $b=0.15$ GeV$^{-2}$ in the HIJING model, the QCD coupling constant $\alpha_s=0.33$, and the screening mass $\mu=3.2$ fm$^{-1}$ to obtain a parton scattering cross section of 1.5 mb in ZPC  for both Pb+Pb collisions at $\sqrt{s_{NN}}=5.02$ TeV and Xe+Xe collisions at $\sqrt{s_{NN}}=5.44$ TeV. The new coalescence parameter $r_{BM}$ is fixed to be 0.61.  These parameters give a better description of the baryon observables, especially the transverse momentum spectra of baryons and antibaryon-to-baryon ratios for $\Xi$ and $\Omega$ in heavy-ion collisions at RHIC and LHC. It is seen that the improved AMPT model almost doesn't affect the flow results and still can reproduce the experimental data of charged pions and kaons elliptic flow at low $p_T$ in Au+Au at $\sqrt{s_{NN}}=$200 GeV and Pb+Pb at $\sqrt{s_{NN}}=$2.76 TeV \cite{He:2017tla}. It should be emphasized that the centrality in the AMPT model is determined by the range of impact parameters in the simulated minimum-bias events. For example, for Xe+Xe collisions at 5.44 TeV the impact parameter ranges 0--2.87 fm and 5.72--7.01 fm correspond to the centralities 0-5\% and 20-30\%, respectively. In our calculations, the hadron interaction is terminated at 200 fm/c.

\begin{figure}[h]
\centerline{
\includegraphics[width=8.5cm]{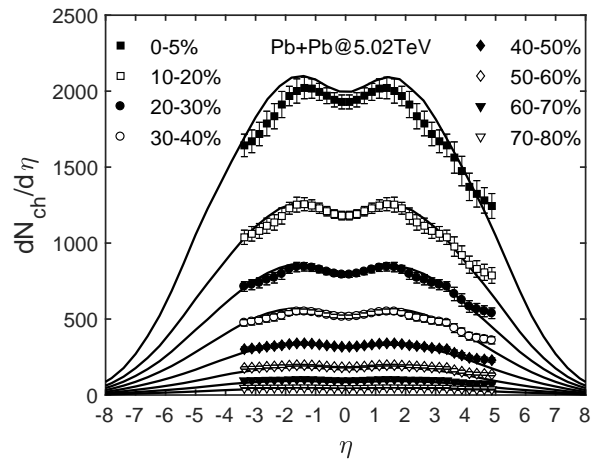}}
\caption{Charged-particle pseudorapidity multiplicity density for eight centrality classes from the improved AMPT model for Pb+Pb collisions at $\sqrt{s_{NN}}=5.02$ TeV. Data  are from the ALICE Collaboration~\cite{Adam:2016ddh}.}
\label{fig1}
\end{figure}

\section{results}\label{results}

In the present section, we show results on the charged-particle pseudorapidity multiplicity density $dN_{ch}/d\eta$, transverse momentum spectra of charged pions, kaons and protons, and their ratios $K/\pi$ and $p/\pi$ for Pb+Pb collisions at $\sqrt{s_{NN}}=5.02$ TeV and Xe+Xe collisions at $\sqrt{s_{NN}}=5.44$ TeV. They are generated from the improved string-melting AMPT model described in Section \ref{ampt}.

\begin{figure}[h]
\centerline{
\includegraphics[width=8.5cm]{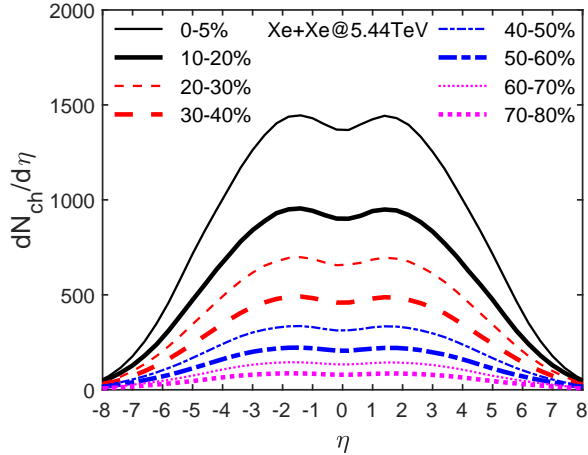}}
\caption{(Color online) Charged-particle pseudorapidity multiplicity density for eight centrality classes from the AMPT model for Xe+Xe collisions at $\sqrt{s_{NN}}=5.44$ TeV.}
\label{dNdeta_xe}
\end{figure}

\subsection{charged-particle pseudorapidity multiplicity density}
Figure \ref{fig1} presents the pseudorapidity multiplicity density of charged particles for eight centrality classes in Pb+Pb collisions at $\sqrt{s_{NN}}=5.02$ TeV with black lines generated from the improved AMPT model. For comparison, we also show the corresponding experimental data from ALICE Collaboration \cite{Adam:2016ddh} by different symbols. As one can see, the results from the improved AMPT model can describe very well the experimental data except that it is slight lower than the experimental data at the forward pseudorapidity $\eta>4$ for the eight centralities. Previously, the new introduced coalescence parameter $r_{BM}$ was fixed by reproducing the proton rapidity distribution $dN/dy$ at midrapidity for central Au+Au collisions at $\sqrt{s_{NN}}=200$ GeV and central Pb+Pb collisions at $\sqrt{s_{NN}}=2.76$ TeV. Our results corroborate that this value still works for higher colliding energy $\sqrt{s_{NN}}=5.02$ TeV and this gives us the confidence to predict the charged-particle pseudorapidity multiplicity density in Xe+Xe collisions at $\sqrt{s_{NN}}=5.44$ TeV using the improved AMPT model with the same parameters selected, see section \ref{ampt}.

We simulate Xe+Xe collisions at $\sqrt{s_{NN}}=5.44$ TeV for eight centrality classes and show the charged-particle pseudorapidity multiplicity density in Fig. \ref{dNdeta_xe}. We notice that the experimental results of the pseudorapidity multiplicity density of charged particles from ALICE Collaboration are presented in Ref. \cite{Acharya:2018hhy}, but  the data have not been available online yet. We hope that the experimental data will still prove the validity of the improved AMPT model in the near future.

\begin{figure*}[t]
\centering
\begin{tabular}{ccc}
\includegraphics[width=0.85\textwidth]{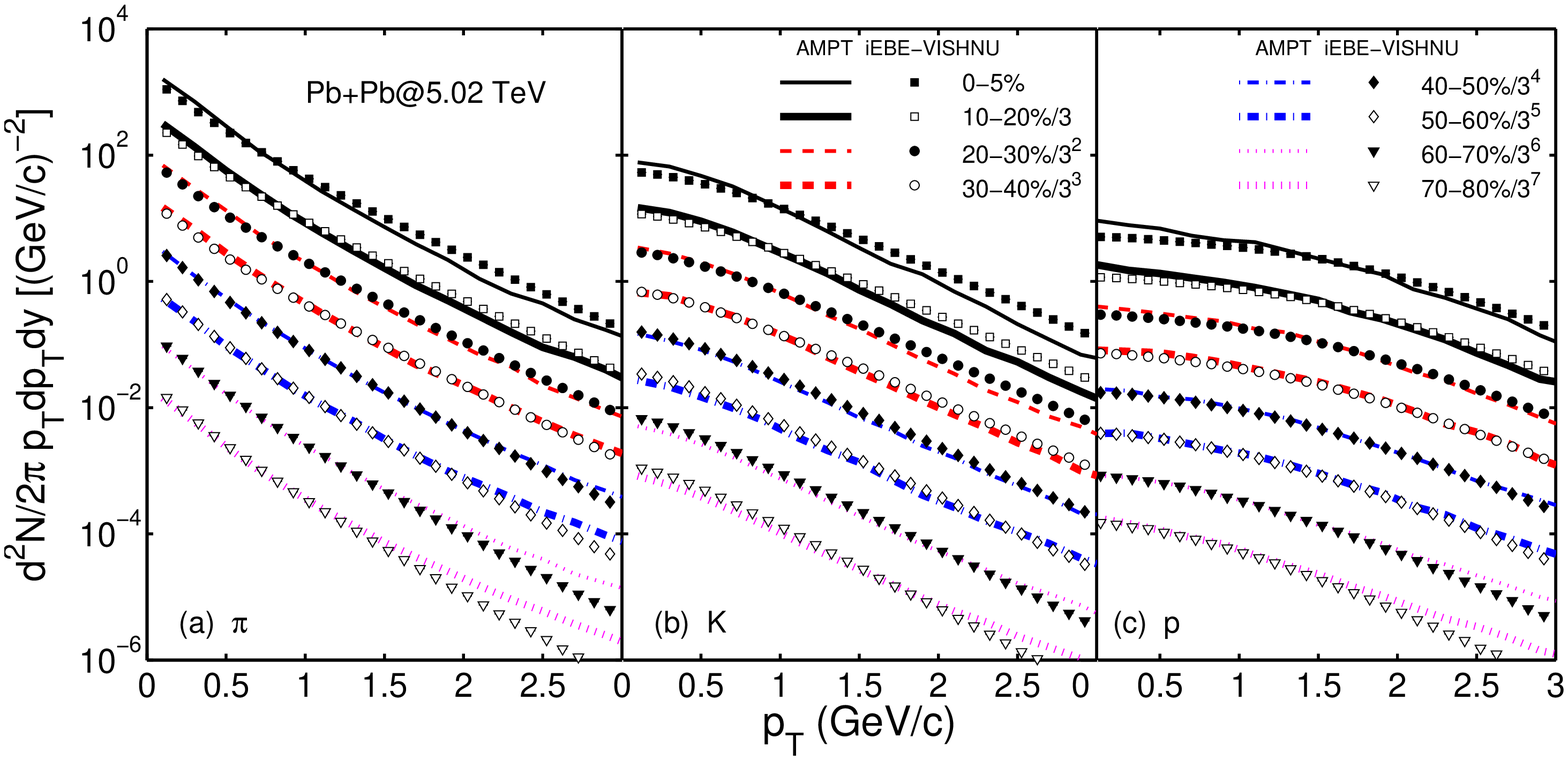}
\end{tabular}
\caption{ (Color online) Transverse momentum spectra of charged pions (left panel), kaons (middle panel) and protons (right panel) for eight centrality classes from the AMPT model for Pb+Pb collisions at $\sqrt{s_{NN}}=5.02$ TeV. The symbols are the predictions of the iEBE-VISHNU model with TRENTo initial condition for Pb+Pb collisions at $\sqrt{s_{NN}}=5.02$ TeV \cite{Zhao:2017yhj}. }\label{spectra_pb}
\end{figure*}

\begin{figure*}[t]
\centering
\begin{tabular}{ccc}
\includegraphics[width=0.85\textwidth]{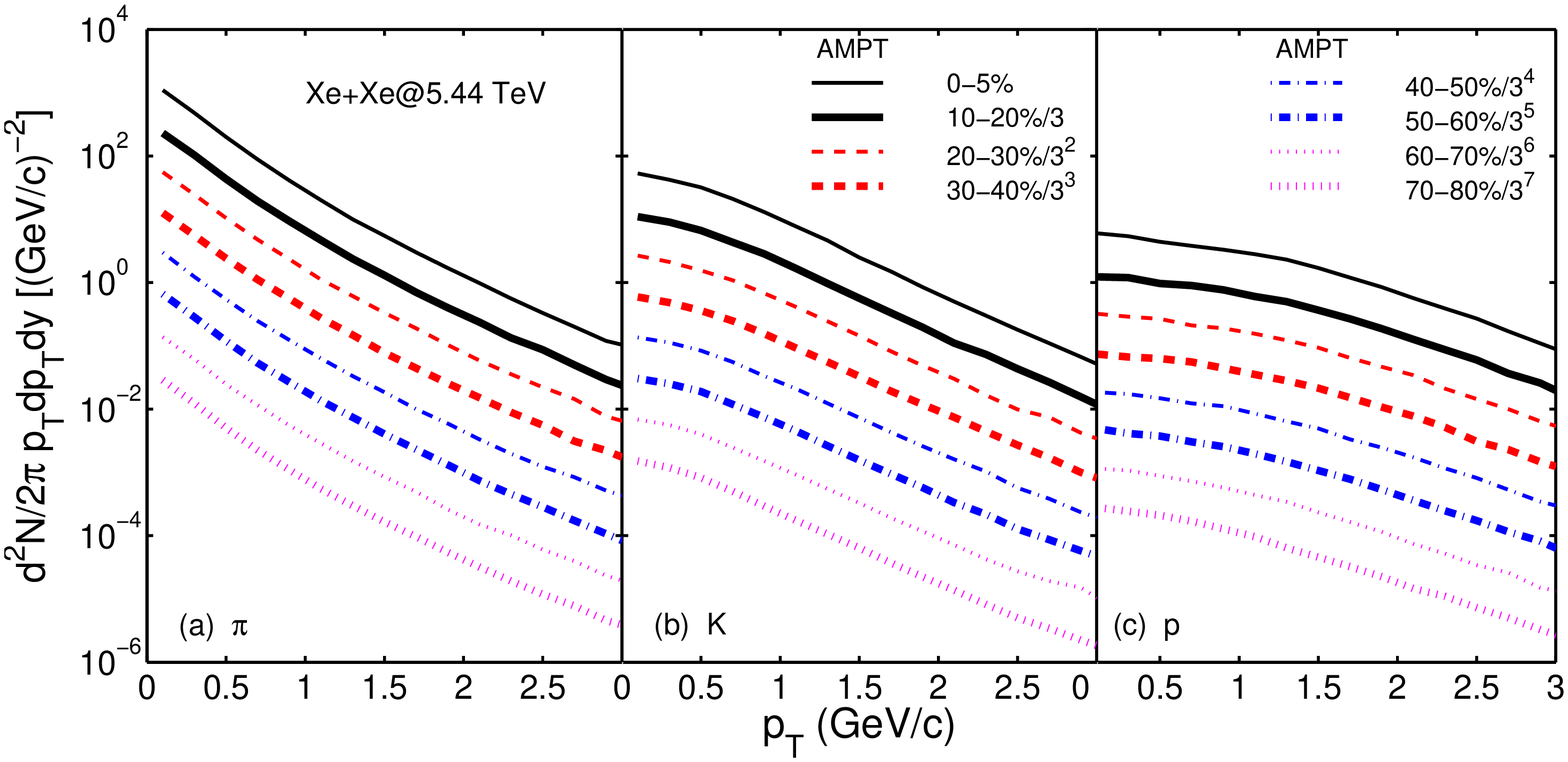}
\end{tabular}
\caption{ (Color online) Transverse momentum spectra of charged pions (left panel), kaons (middle panel) and protons (right panel) for eight centrality classes from the AMPT model for Xe+Xe collisions at $\sqrt{s_{NN}}=5.44$ TeV.}
\label{spectra_xe}
\end{figure*}

\subsection{Transverse momentum spectra}

Figure \ref{spectra_pb} shows the transverse momentum spectra of charged pions, kaons and protons from the improved AMPT model for various centralities for Pb+Pb collisions at $\sqrt{s_{NN}}=5.02$ TeV. The experimental data for the $p_T$ spectra of these particles are not available yet, so we cannot compare our results with the data directly. The hydrodynamics model is a very convincing tool to study the physics at low transverse momentum region. One such standard model is iEBE-VISHNU hybrid model \cite{Shen:2014vra}, which is an event-by-event version of the VISHNU hybrid model. It combines (2+1)-d viscous hydrodynamics VISH2+1 \cite{Song:2007fn, Song:2009gc} to describe the expansion of the QGP fireball with a hadron cascade model (UrQMD) \cite{Bass:1998ca, Bleicher:1999xi} to simulate the evolution of the hadronic matter. iEBE-VISHNU has been successfully used to describe various soft physics at RHIC and LHC in the past few years \cite{Zhao:2017yhj, Zhu:2016puf, Xu:2016hmp}. In Ref. \cite{Zhao:2017yhj}, the authors showed that iEBE-VISHNU can reproduce the transverse momentum spectra of charged pions, kaons and protons produced in Pb+Pb collisions at $\sqrt{s_{NN}}=2.76$ TeV and predicted the same observables for Pb+Pb collisions at $\sqrt{s_{NN}}=5.02$ TeV. As a reference to our results from the improved AMPT model, we put the prediction from iEBE-VISHNU model with TRENTo initial condition in Fig. \ref{spectra_pb}, shown by different symbols for different centralities. As we can see, the results from the two models are very similar. For pions, the results from AMPT model at central collisions are a little bit lower than those from iEBE-VISHNU hybrid model at $p_T>2$ GeV/c, while the situation is opposite for the very peripheral collisions. But the two models give the same results for the semi-central collisions. The similar behaviors can be found for kaons as well as protons in Fig. \ref{spectra_pb}(b) and (c), respectively. These results empower us to investigate the transverse momentum spectra of identified particles produced in  Xe+Xe collisions at $\sqrt{s_{NN}}=5.44$ TeV.

We show the transverse momentum spectra of charged pions, kaons and protons for Xe+Xe collisions at $\sqrt{s_{NN}}=5.44$ TeV in Fig. \ref{spectra_xe}. Since the colliding energies for the two systems, i.e., Pb+Pb and Xe+Xe, are so close, we expect that the general behaviors for charged pions, kaons and protons produced in the two colliding systems are very similar and indeed they are. The spectra get flatter with the increasing mass of the identified particle of interest, e.g. from pions to protons. On the other hand, the mass number of the xenon nuclei is roughly halfway that of lead nuclei, so the spectra of charged pions, kaons and protons in Pb+Pb collisions are higher  for every centrality than the ones at Xe+Xe collisions which simply tells us that larger final multiplicity is produced in Pb+Pb collisions at $\sqrt{s_{NN}}=5.02$ TeV than the one in Xe+Xe collisions at $\sqrt{s_{NN}}=5.44$ TeV. This can also be seen from the charged particle pseudorapidity multiplicity density in Figs. \ref{fig1} and \ref{dNdeta_xe}.

\subsection{particle ratios}

\begin{figure}[h]
\centering
\begin{tabular}{ccc}
\includegraphics[width=8.5cm]{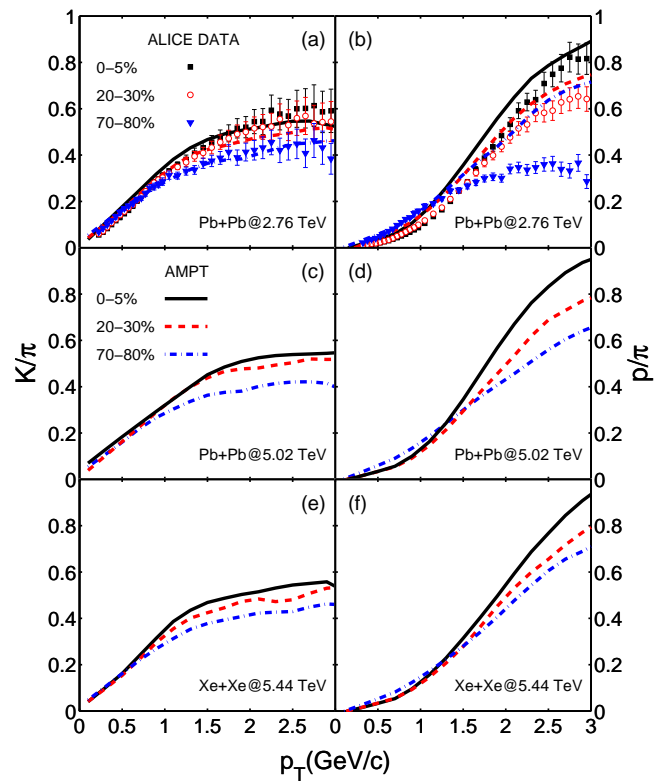}
\end{tabular}
\caption{ (Color online) $K/\pi$ and $p/\pi$ ratios for central (0-5\%), mid-central (20-30\%) and peripheral (70-80\%) Pb+Pb collisions at $\sqrt{s_{NN}}=2.76$ TeV (top panels) and $\sqrt{s_{NN}}=5.02$ TeV (middle panels) and Xe+Xe collisions at $\sqrt{s_{NN}}=5.44$ TeV (bottom panels). The experimental data shown in the top panels are from Ref. \cite{Abelev:2013vea}.}
\label{ratio}
\end{figure}

Figure \ref{ratio} shows the charged particle ratios  $K/\pi$ and $p/\pi$ for the most central (0-5\%, solid lines), mid-central (20-30\%, dashed lines) and peripheral (70-80\%, dash-dotted lines) centrality bins in Pb+Pb collisions at $\sqrt{s_{NN}}=2.76$ TeV (top panels) and 5.02 TeV (middle panels), and  Xe+Xe collisions at $\sqrt{s_{NN}}=5.44$ TeV (bottom panels) from the improved AMPT model. Also shown in the top two panels are the experimental data for $K/\pi$ and $p/\pi$ from the ALICE Collaboaration \cite{Adam:2015kca} for Pb+Pb collisions at $\sqrt{s_{NN}}=2.76$ TeV. Our result reproduces very well the experimentally measured $K/\pi$ in panel (a). For $p/\pi$, the improved AMPT model reasonably describes the experimental data for the two centralities 0-5\% and 20-30\%, which is consistent with the result in Ref. \cite{He:2017tla}, while it overestimates the ratio above 1.5 GeV/c for 70-80\%. It means that the improved AMPT model can describe the baryon production at central and semi-central collisions, but it still doesn't work well for peripheral collisions in the heavy ion collisions. It is out of the scope of this paper to improve the AMPT model for the peripheral collisions.

Figure \ref{ratio}(c) and (e) show the AMPT predictions for the ratio of $K/\pi$ in Pb+Pb collisions at $\sqrt{s_{NN}}=5.02$ TeV and Xe+Xe collisions at $\sqrt{s_{NN}}=5.44$ TeV, respectively. According to the results of Pb+Pb at $\sqrt{s_{NN}}=2.76$ TeV, we expect that our predictions for the particle ratios should be able to describe the coming experimental data quantitatively for Pb+Pb collisions at $\sqrt{s_{NN}}=5.02$ TeV and  Xe+Xe collisions at $\sqrt{s_{NN}}=5.44$ TeV except the $p/\pi$ ratio at peripheral collisions. Since the colliding energies are so close, the particle ratios for the two systems have some properties in common. First of all, the $K/\pi$ ratios are essentially identical for all centrality classes in panels (c) and (e), and very similar to those for Pb+Pb collisions at $\sqrt{s_{NN}}=2.76$ TeV, which may indicate that the mechanism for strange production in Pb+Pb and Xe+Xe collisions is the same. Secondly, since the peripheral collisions in A+A are more similar to p+p collisions, more strange quarks should be produced at higher colliding energies. Therefore, the $K/\pi$ ratio for the centrality 70-80\% in Xe+Xe collisions is a little bit higher than that in Pb+Pb collisions. The ratios $p/\pi$ for Pb+Pb collisions at $\sqrt{s_{NN}}=5.02$ TeV and Xe+Xe collisions at $\sqrt{s_{NN}}=5.44$ TeV are established in Fig. \ref{ratio}(d) and (f), which rise steeply at low $p_T$. The increase is faster in central collisions than in peripheral ones. This is conjectured to be attributed to the parton recombination mechanism of hadronization, which gives rise to a significant enhancement of baryon yields relative to meson yields in heavy ion collisions \cite{vg, rf, hy1}. For the Pb+Pb and Xe+Xe collisions considered here, thousands of soft hadrons and multiple hard jets are created. Minijets that are copiously produced at intermediate $p_T$ can fragment into soft partons with multiplicities so high that their effects on the hadronization of all partons created in the soft sector cannot be ignored. Therefore, it leads to large $p/\pi$ ratio at intermediate $p_T$ region.

\section{summary}\label{summary}

Using the improved string-melting AMPT model, we have studied and predicted the charged-particle pseudorapidity multiplicity density, $p_T$ spectra for charged pions, kaons and protons and their ratios from central to peripheral collisions in Pb+Pb collisions at $\sqrt{s_{NN}}=5.02$ TeV and Xe+Xe collisions at $\sqrt{s_{NN}}=5.44$ TeV. More specifically, we have calculated the charged-particle pseudorapidity multiplicity density in Pb+Pb collisions at $\sqrt{s_{NN}}=5.02$ TeV, which are seen to agree with the experimental data from the ALICE Collaboration, while the calculated transverse momentum spectra of charged pions, kaons and protons are consistent with the results from the iEBE-VISHNU hybrid model with TRENTo initial condition. It indicates that the two models give the similar space-time evolution of the hot dense matter created in heavy-ion collisions. With the same parameters, we predict the centrality dependence of charged-particle pseudorapidity multiplicity density and transverse momentum spectra of charged pions, kaons and protons in Xe+Xe collisions at $\sqrt{s_{NN}}=5.44$ TeV. Furthermore, the charged particle ratios of $K/\pi$ and $p/\pi$ for Pb+Pb collisions at $\sqrt{s_{NN}}=5.02$ TeV and  Xe+Xe collisions at $\sqrt{s_{NN}}=5.44$ TeV are also predicted. As a reference, the ratios $K/\pi$ and $p/\pi$ from Pb+Pb collisions at $\sqrt{s_{NN}}=2.76$ TeV are also calculated and compared with the experimental data. These predictions presented in this  paper will test the predictive power of the improved AMPT model when the experimental data are available.

\section*{Acknowledgements}
The authors would like to thank W.B. Zhao for giving us the simulation results of Pb+Pb collisions at $\sqrt{s_{NN}}=5.02$ TeV from iEBE-VISHNU model and Z. W. Lin for providing the code of improved AMPT model. This work was supported by the NSFC of China under Grant No. 11205106.

\end{document}